\documentstyle[aps,multicol,psfig]{revtex}

\draft

\begin{document}

\title{Diffusion regimes in L\'evy flights with trapping}
\author{Alexei Vazquez$^1$, Oscar Sotolongo-Costa$^{1,2}$ and Francois Brouers$^3$}

\address{$^1$ Departamento de Fisica Teorica Universidad de La Habana, Habana 10400, Cuba.}

\address{$^2$ Departamento de Fisica Fundamental, L.C.T.D.I, Fac. Ciencias, UNED, Madrid, Spain}

\address{$^3$ Institute of Physics, Li\'ege University, B5, 4000 Li\'{e}ge, Belgium}

\date{\today}

\maketitle

\begin{abstract}
The diffusion of a walk in the presence of traps is investigated. Different
diffusion regimes are obtained considering the magnitude of the fluctuations
in waiting times and jump distances. A constant velocity during the jump
motion is assumed to avoid the divergence of the mean squared displacement.
Using the limit theorems of the theory of L\'{e}vy stable distributions we
have provided a characterization of the different diffusion regimes.\newline

\end{abstract}


\begin{multicols}{2}
 
\section{Introduction}

There are many physical situations where diffusion takes place under the
presence of traps, trapping diffusion. Examples are found in electronic
conduction in amorphous semiconductors and quasicrystals \cite
{semiconductors}, atomic diffusion in glass like materials \cite{glasses},
tracer diffusion in living polymers \cite{polymers}, and more \cite
{bouchoud,montroll}.The existence of traps is generally modeled through a
probability density of waiting times between successive steps in the walk,
continuous-time random walks (CTRW) \cite{montroll}. The theory of CTRW has
been extensively developed in the literature, using the generating function
methods \cite{montroll} or simple statistical reasoning \cite{bouchoud}. The
existence of a wide distribution of waiting times leads to a subdiffusive
regime where the mean squared displacement grows slower than time. On the
other hand, there are physical situations where the time between successive
jumps may be considered constant, but the distribution of jump distances is
wide \cite{mandelbrot,shlesinger,bouchoud,montroll}, (L\'{e}vy flights) \cite
{mandelbrot}.

L\'{e}vy flights are also observed in a large variety of phenomena, for
instance in chaotic dynamical systems \cite{shlesinger,zumofen}, turbulent 
\cite{hayot,solomon} flow and self-gravitating systems \cite{antoni}. The
existence of a wide distribution of jump distances leads to a superdiffusive
regime, where the mean squared displacement grows faster than time.
Moreover, L\'{e}vy flights generates fractal structures in space \cite
{mandelbrot}. Nevertheless, more complex behaviors are expected in systems
where both the distribution of waiting times and jump distances are wide.
For instance, Schulz \cite{schulz} studied an anomalous Drude model for
transport properties of quasicrystals. He assumed that the walks move with
an anomalous speed $v_\sigma \sim t^\sigma $ between collisions, and a power
law $p(\tau )\sim \tau ^{-1-\mu }$ distribution of time between collisions.
In this way, he obtained the phase diagram ($\mu ,\sigma $) of the system,
which is divided in different regions. While the anomalous velocity may
mimic the existence of traps, it is introduced artificially and cannot be
derived from a Hamiltonian. In \cite{kbs}, Klafter {\em et al }introduced a
stochastic description of anomalous transport phenomena and found different
behavior of the mean square displacement of CTRW. Zumofen and Klafter \cite
{zumofen} studied a one-dimensional map which exhibits intermittent chaotic
behavior with coexisting laminar and localized phases, and analyzed them in
terms of L\'{e}vy statistics.

A more consistent approach was presented by Klafter and Zumofen \cite
{klafter}. They studied the diffusion in a Hamiltonian system in terms of
the CTRW formulation, and considered wide distributions of waiting times and
jump distances. However, they did not provide a complete characterization of
the different diffusion regimes that may be obtained. In a recent letter
Fogedby \cite{fogedby} considered Le\'{v}y flights in the presence of a
quenched isotropic random force field studying the interplay between the
built in superdiffusive behavior of the Levy flights and the pinning effect
of the random environment.

In the present work we study the behavior of a random walk in the presence
of traps. Disorder is introduced in an annealed way, through power law
distributions of waiting times $p(\tau )\sim \tau ^{-1-\alpha _w}$ and jump
distances $p(x)\sim x^{-1-\alpha _x}$. We use simple statistical reasoning
close to the approach developed by Bouchaud and Georges \cite{bouchoud}
instead of that of the work by Klafter and Zumofen \cite{klafter}, which use
the CTRW formulation. The present formalism contains as a fundamental tool
the theory of L\'{e}vy stable distributions, introduced by L\'{e}vy \cite
{levy} and developed by other authors \cite{feller,gnedenko}.

\section{The model}

We consider a random walk on a lattice, such that the particle has to wait
for a time $\tau _w$ on each site before performing the next jump. The
waiting time is a random variable independently chosen at each new jump
according to a distribution $p(\tau _w)$. We also assume that the waiting
time is not correlated to the length of the jump $x$, which is distributed
according to $p(x)$. The diffusion process will be characterized by the
scattering function $F(k,t)$, the Fourier transform of the diffusion front.
Other properties like the diffusion front and the mean squared displacement
can be derived from this function. For instance, the mean squared
displacement is given by 
\begin{equation}
\langle x^2(t)\rangle =-\frac{\partial ^2F}{\partial k^2}\biggl|_{k=0}.
\label{eq:0}
\end{equation}
Let $N$ be the number of steps performed by a walker during time $t$. $N$
is, in general, a random variable which depends on the duration of the jumps
and waiting times. The scattering function can thus be expressed as a sum
over all possible jumps during time $t$%
\begin{equation}
F(k,t)=\int dNF(k,N)P(N,t),  \label{eq:1}
\end{equation}
where $F(k,t)$ is the scattering function of the same problem, but
considering regular duration of the jumps and no waiting time and $P(N,t)$
stands for the probability distribution of $N$ jumps at a fixed time $t$.

\subsection{Mean squared displacement after $N$ steps}

The total displacement after $N$ steps is given by 
\begin{equation}
X_N=\sum_{i=1}^Nx_i.  \label{eq:2}
\end{equation}
In the right hand side we have a sum of mutually independent random
variables with the common distribution $p(x)$, with zero mean. The limit
distribution for large $N$ will be a stable L\'{e}vy distribution \cite
{feller,gnedenko}, i.e. 
\begin{equation}
X_N\dot{=}l^{*}N^{1/\alpha _x}u  \label{eq:3}
\end{equation}
where $\dot{=}$ denotes that random variables in both sides have the same
distribution, $l^{*}$ is a characteristic value and $u$ follows the
symmetric L\'{e}vy distribution $L_{\alpha _x0}(u)$. The canonical (Fourier
transform) representation of L\'{e}vy stable laws is (for $\alpha \neq 1$) 
\begin{equation}
\text{FT}[L_{\alpha \beta }](k)=\exp \biggl[-|k|^\alpha \biggl(1+\frac
k{|k|}i\beta \tan \alpha \frac \pi 2\biggr)\biggr],  \label{eq:4}
\end{equation}
where $\alpha $ and $\beta $ are real numbers defined in the intervals $%
0<\alpha \leq 2$ and $-1\leq \beta \leq 1$. The case $\alpha =2$ and $\beta
=0$ corresponds with the Gaussian distribution, which decays faster than any
power law for large arguments. On the contrary, all L\'{e}vy distributions,
except the Gaussian, have the asymptotic behavior for $u\gg 1$ \cite
{feller,gnedenko} 
\begin{equation}
L_{\alpha \beta }(u)\sim u^{-1-\alpha }.  \label{eq:5}
\end{equation}
Then, from eqs. (\ref{eq:3}) and (\ref{eq:4}) it follows that 
\begin{equation}
F(k,N)=\exp [-(|k|l^{*})^{\alpha _x}N],  \label{eq:6}
\end{equation}
If $p(x)$ has finite variance $\sigma $ then $l^{*}=\sigma $ and $\alpha
_x=2 $. If $p(x)\approx l_0^{\alpha _x}|x|^{-1-\mu }$, with $0<\mu <2$ then $%
l^{*}\sim l_0$ and $\alpha _x=\mu $.

\begin{figure}\narrowtext
\centerline{\psfig{figure=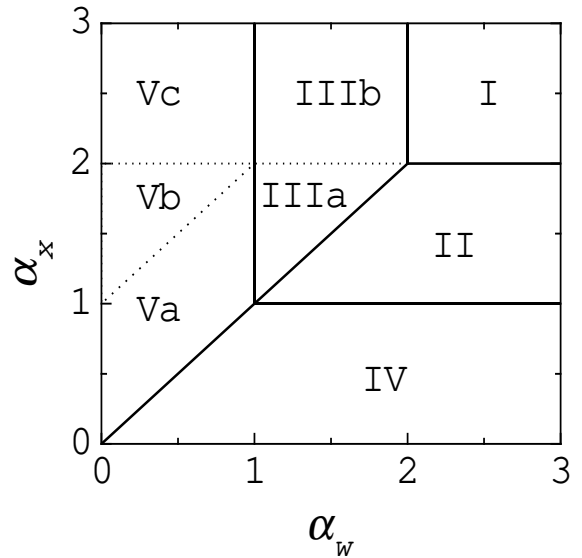,width=3in,height=3in}}
\caption{The phase diagram ($\alpha _w$ , $\alpha _x$). The behavior in the
different regions is as follows: I.- normal diffusion; II.- LTT with
exponent $\alpha _x$ and superdiffusion; III.- LTT with exponent $\alpha _w$
a) superdiffusive, b) normal diffusion; IV.- LTT with exponent $\alpha _x$
and ballistic motion; V.- LTT with exponent $\alpha _w$, a) superdiffusive,
b) and c) subdiffusive. See text for a detailed description.}
\label{fig:1}
\end{figure}

\subsection{Number of steps after time $t$}

On the other hand, the number of steps after time $t$ is given by 
\begin{equation}
t=\sum_{i=1}^N\tau _{wi}+\sum_{i=1}^N\tau _{xi}.  \label{eq:7}
\end{equation}
where $\tau _{xi}$ are the duration of the jumps. If we assume that during
the jump motion the walker moves continuously at a constant velocity $v$ and
changes directions at random then $\tau _x=v^{-1}|x|$ and $p(\tau
_x)=2vp(|x|/v)$. In the right hand side of eq. (\ref{eq:7}) we have two sums
of independent random variables, with common distribution $p(\tau _w)$ and $%
p(\tau _x)$, respectively. The limit distributions for large $N$ will follow
L\'{e}vy distributions \cite{feller,gnedenko}, i.e. 
\begin{equation}
t=\tau N+\tau _w^{*}N^{1/\alpha _w}u_1+\tau _x^{*}N^{1/\alpha _x}u_2.
\label{eq:8}
\end{equation}
where $u_1$ and $u_2$ follows the L\'{e}vy distribution $L_{\alpha _w1}(u_1)$
and $L_{\alpha _x1}(u_2)$, respectively. The first term in the right hand
side appears only if $p(\tau _w)$ or $p(\tau _x)$ have finite mean, and $%
\tau $ is given by the sum of the finite means. If $p(\tau _w)$ has finite
variance $\sigma $ then $\tau _w^{*}=\sigma $ and $\alpha _w=2$, while if $%
p(\tau _w)\approx \tau _0^{\alpha _w}\tau _w^{-1-\mu }$ ($0<\mu <2$) then $%
\tau _w^{*}\sim \tau _0$ and $\alpha _w=\mu $. If $p(\tau _x)$ has finite
variance $\sigma $ then $\tau _x^{*}=\sigma $ and $\alpha _x=2$, while if $%
p(\tau _x)\approx \tau _0^{\alpha _x}\tau _x^{-1-\mu }$ ($0<\mu <2$) then $%
\tau _x^{*}\sim \tau _0$ and $\alpha _x=\mu $.From eqs. (\ref{eq:1}), (\ref
{eq:6}) and (\ref{eq:8}) it follows that 
\begin{equation}
F(k,t)=\int \int du_1du_2L_{\alpha _w1}(u_1)L_{\alpha _x1}(u_2)\exp
[-(kl^{*})^{\alpha _x}N].  \label{eq:9}
\end{equation}
where the functional dependence of $N$ with $u_1$, $u_2$ and $t$ is
determined from eq. (\ref{eq:8}). Next we analyze the behavior of the
scattering function, defined trough this expression, for different values of 
$\alpha _x$ and $\alpha _w$. With this purpose we have divided the
corresponding phase diagram in five regions, as it is illustrated in fig. 1.

\section{\protect\bigskip Results}

In region I both the distribution of waiting times and jump distances have
finite variance. Hence, the random variables $u_1$ and $u_2$ will follow a
Gaussian distribution. For large $N$, we can therefore neglect the last two
terms in the right hand side of eq. (\ref{eq:8}), obtaining $N\approx t/\tau 
$. Then eq. (\ref{eq:9}) reduces to 
\begin{equation}
F(k,t)=\exp [-(|k|l^{*})^2t/\tau ].  \label{eq:10}
\end{equation}
Moreover, using eq. (\ref{eq:0}) one obtains 
\begin{equation}
\langle x^2(t)\rangle \sim t.  \label{eq:11}
\end{equation}
We thus find normal or classical diffusion in this region: the scattering
function decays exponentially with time and the mean squared displacement
grows proportional to time.

In region IV the distribution of jump distances is quite wide, and even
wider than the distribution of waiting times. Hence, it is expected that the
third term in eq. (\ref{eq:8}) gives the major contribution, thus obtaining $%
N\approx (t/\tau _x^{*})^{\alpha _x}u_2^{-\alpha _x}$. Substituting this
result in eq. (\ref{eq:9}), and expanding the exponential inside the
integral, one obtains 
\begin{equation}
F(k,t)=\sum_{n=0}^\infty \frac{(-1)^n}{\Gamma (1+n\alpha _x)}(\frac t{\tau
_k})^{n\alpha _x},  \label{eq:12}
\end{equation}
where $\tau _k=\tau _x^{*}/|k|l^{*}$. This series has an infinite radius of
convergence and, therefore, can be taken as a series expansion for the
scattering function in this region. For small times ($t\ll \tau _k$) the
scattering function follows a stretched exponential decay, with stretched
exponent $\alpha _x$. On the contrary, for $t\gg \tau _k$ the relaxation
becomes slower than an exponential. Taking $L_{\alpha _x\ 1}(u_1)\sim
u_1^{-1-\alpha _x}$, from eq. (\ref{eq:9}) it follows that $F(k,t)\sim
t^{-\alpha _x}$. Hence, for long times the scattering function follows a
long time tail (LTT), with an exponent $\alpha _x$ smaller than one. In this
case, the mean squared displacement determined from eqs. (\ref{eq:0}) and (%
\ref{eq:12}) is not finite. This is a consequence of the divergence of the
second moment of the distribution of jump distances. Nevertheless, the total
displacement at time $t$ cannot be larger than $vt$ and, therefore, there is
a cutoff $k_c\sim t^{-1}$ for small values of $k$. Thus, to avoid the
divergence of the mean squared displacement we evaluate eq. (\ref{eq:0}) in $%
k=k_c$ instead of $k=0$. In this way we obtain 
\begin{equation}
\begin{array}{ll}
\langle x^2(t)\rangle \sim t^2, & \text{in IV}.
\end{array}
\label{eq:13}
\end{equation}
The motion of the walk is in this case of ballistic type. The behavior in
this region has been investigated by different authors, which consider
random walk motion due to a periodic potential \cite{klafter}.

In region V we can use a similar reasoning as in region IV, but in this case
the dominant term will be the second one, associated to huge fluctuations in
the waiting time. In this way we obtain 
\begin{equation}
F(k,t)=\sum_{n=0}^\infty \frac{(-1)^n}{\Gamma (1+n\alpha _w)}(\frac t{\tau
_k})^{n\alpha _w},  \label{eq:14}
\end{equation}
where $\tau _k=\tau _w^{*}(|k|l^{*})^{-\alpha _x/\alpha _w}$. This
expression is similar to eq. (\ref{eq:12}). Besides, the asymptotic
behaviors for short and long times are the same, replacing $\alpha _x$ by $%
\alpha _w$. The main difference is observed in the $k$-dependence of the
relaxation time $\tau _k$, which is manifested in the temporal dependence of
the mean squared displacement, 
\begin{equation}
\langle x^2(t)\rangle \sim \left\{ 
\begin{array}{ll}
t^{2-\alpha _x+\alpha _w}, & \text{in Va and Vb}; \\ 
t^{\alpha _w} & \text{in Vc}.
\end{array}
\right.  \label{eq:15}
\end{equation}
The difference between region Va and Vb is that, in the former the mean
squared displacement grows faster than time and, therefore, the system is in
a superdiffusive regime, while in the second one it grows slower than time
and the system is in a subdiffusive regime. In region Vc there is also a
subdiffusive regime, which has been investigated by different authors, using
the generating function formulation \cite{montroll} or simple statistical
reasoning \cite{bouchoud}. However, in region Vc $\alpha _x>2$ and,
therefore, the spatial trajectories of the walk will not be self-similar as
in regions Va and Vb.

In region II and III the fluctuations in jump distances and waiting times,
respectively, are still L\'{e}vy type, but are no so strong as in regions IV
and V, respectively. In these cases one cannot neglect the first term in the
right hand side of eq. (\ref{eq:8}) and, hence, there is some reminiscent of
normal diffusion behavior.

In region II the distribution of jump distances is wider than the
distribution of waiting times, and both have finite mean. We thus expect
that the third term in the right hand side of eq. (\ref{eq:8}) gives the
major contribution to the fluctuations in $N$, and the second one may be
neglected. Even with this simplification we cannot solve eq. (\ref{eq:8})
analytically, however the following asymptotic behaviors are obtained 
\begin{equation}
N\approx \left\{ 
\begin{array}{ll}
t/\tau & u\ll u_c(t), \\ 
(t/\tau _x^{*})^{\alpha _x} & u\gg u_c(t);
\end{array}
\right. ,  \label{eq:16}
\end{equation}
where $u_c(t)\sim (t/\tau )^{\alpha _x}$. For small times, substituting eq. (%
\ref{eq:16}) in eq. (\ref{eq:9}) one obtains 
\begin{equation}
F(k,t)=\exp [-(kl^{*})^{\alpha _x}t/\tau ].  \label{eq:17}
\end{equation}
The relaxation for small times is the exponential like in the normal
diffusion case, region I. However the mean squared displacement does not
grow linearly with time, 
\begin{equation}
\begin{array}{ll}
\langle x^2(t)\rangle \sim t^{3-\alpha _x}, & \text{in II},
\end{array}
\label{eq:18}
\end{equation}
but is characteristic of a superdiffusion regime. For long times, from eqs. (%
\ref{eq:9}) and eq. (\ref{eq:16}), and the asymptotic expansion for large
arguments of L\'{e}vy distributions in eq. (\ref{eq:5}), it follows that $%
F(k,t)\sim t^{-\alpha _x}$. For long times the scattering function decays
slower than an exponential following a power tail, like in region IV, but
with a larger exponent.

In region III we can use a similar reasoning as in region II, but now
neglecting the fluctuations in the jump distances in relation to the
fluctuations in the waiting times. Again, even with this simplification we
cannot solve eq. (\ref{eq:8}) analytically, however the following asymptotic
behaviors are obtained 
\begin{equation}
N\approx \left\{ 
\begin{array}{ll}
t/\tau & u\ll u_c(t), \\ 
(t/\tau _w^{*})^{\alpha _w} & u\gg u_c(t);
\end{array}
\right.  \label{eq:19}
\end{equation}
where now $u_c(t)\sim (t/\tau )^{\alpha _w}$. For small times, substituting
eq. (\ref{eq:19}) in eq. (\ref{eq:9}) one obtains an exponential relaxation
as in eq. (\ref{eq:17}). However, the characteristic exponent $\alpha _x$
can be now larger than two leading to different behaviors for the mean
squared displacement, 
\begin{equation}
\langle x^2(t)\rangle \sim \left\{ 
\begin{array}{ll}
t^{3-\alpha _x}, & \text{in IIIa}; \\ 
t & \text{in IIIb}.
\end{array}
\right.  \label{eq:20}
\end{equation}
Hence, in subregions IIIa and IIIb we found superdiffusion and normal
diffusion behavior, respectively. For long times, from eqs. (\ref{eq:9}) and
eq. (\ref{eq:19}), and the asymptotic expansion for large arguments of
L\'{e}vy distributions in eq. (\ref{eq:5}), it follows that $F(k,t)\sim
t^{-\alpha _w}$. For long times the scattering function decays slower than
an exponential following a power tail, like in region V, but with a larger
exponent.

\section{Summary an conclusions}

In summary we have investigated the diffusion behavior in the whole plane ($%
\alpha _w,\alpha _x$), which has been divided in five regions considering
the magnitude of the fluctuations in waiting times and jump distances. A
constant velocity during the jump motion was assumed to avoid the divergence
of the mean squared displacement. Using as a fundamental tool the limit
theorems of the theory of L\'{e}vy distributions we have provided a
characterization of the different diffusion regimes.

We conclude that the diffusion behavior cannot be just classified as normal
diffusion, superdiffusion and subdiffusion. This classification only takes
into account the temporal dependence of the scattering function for short
times, while the long time decay may be different. For instance, in regions
II and IIIa we have similar superdiffusive regimes, but the long time
behavior is determined by the distribution of jump distances, in the first
case, and by the distribution of waiting times in the second one.

\section*{Acknowledgments}

This work was partially supported by the {\em Alma Mater} prize,given by The
University of Havana. F. B. is grateful to the TWAS visiting professorship
program. O. S.. is grateful to UNED for kind hospitality during his sabbatic
leave from Havana University. We want to express our gratitude to E. Barkai
for helpful comments and suggestions.

\end{multicols}

\end{document}